# Calculating the Likelihood Ratio for Multiple Pieces of Evidence

**Norman Fenton**[1] **and Martin Neil**

**9 June 2021**

## Abstract

When presenting forensic evidence, such as a DNA match, experts often use the Likelihood ratio (LR) to explain the impact of evidence. The LR measures the probative value of the evidence with respect to a single hypothesis such as "DNA comes from the suspect", and is defined as the probability of the evidence if the hypothesis is true divided by the probability of the evidence if the hypothesis is false. The LR is a valid measure of probative value because, by Bayes Theorem, the higher the LR is, the more our belief in the probability the hypothesis is true increases after observing the evidence. The LR is popular because it measures the probative value of evidence without having to make any explicit assumptions about the prior probability of the hypothesis. However, whereas the LR can in principle be easily calculated for a distinct single piece of evidence that relates directly to a specific hypothesis, in most realistic situations 'the evidence' is made up of multiple dependent components that impact multiple different hypotheses. In such situations the LR cannot be calculated. However, once the multiple pieces of evidence and hypotheses are modelled as a causal Bayesian network (BN), any relevant LR can be automatically derived using any BN software application.

[1] Corresponding author: n.fenton@qmul.ac.uk, Twitter: @ProfNFenton

# 1 Introduction: Legal reasoning and Bayes

When a person is accused of committing an offence (civil or criminal) any outside observer with an interest in the case (be it the police, a lawyer, jury member or just a member of the public) may consider whether the defendant is guilty as charged and with what probability or belief. Before they are made aware of any evidence for or against the defendant, these initial personal beliefs will likely be based on relevant background information available to them. As evidence is presented, they will update their beliefs about whether or not the person is guilty based on whether this evidence supports the prosecution or defence arguments. It is very rare for any one piece of evidence to 'prove' guilt or innocence. An independent eyewitness stating that they saw the defendant commit the offence is not proof of guilt, and an independent eyewitness providing an alibi is not proof of innocence; in either case there is uncertainty about both the reliability of the witness and accuracy of their judgement. However, in the former case the evidence 'supports' guilt in the sense that it ***increases*** our belief in guilt, while in the latter case it 'supports' innocence because it ***reduces*** our belief in guilt.

This legal paradigm of updating our belief in an unknown hypothesis in the light of new evidence is a good match for Bayesian probabilistic reasoning.

Formally, with Bayes there is some uncertain hypothesis *H* which, for simplicity, we will assume is either true or false. For example, *H* may be "defendant is guilty as charged" in which case the true state can be denoted *Hp* (the 'prosecution hypothesis') and the false state denoted *Hd* ("defence hypothesis"). We start with our ***prior probability*** of *Hp* which we write as P(*Hp*) so that we know from the axioms of probability theory that P(*Hd*)=1 - P(*Hp*). When we receive evidence *E* we revise our belief in *Hp* to form our ***posterior probability*** of *Hp* which is written as:

P(*Hp* given *E*), or more formally and concisely as P(*Hp* | *E*)

which is the probability of *Hp* given that *E* is true.

Provided we know the prior probability P(*Hp*), Bayes theorem (as we will show) enables us to calculate the posterior probability P(*Hp* | *E*) in terms of a related but different probability, namely P(*E* | *Hp*), i.e. the probability of the evidence given *Hp*. Consider, for example, the following scenario:

> A crime has been committed and a fragment of DNA (believed to be from the person who committed the crime) is recovered from the crime scene. Forensic experts identity a partial DNA profile from the crime scene fragment and determine that it occurs in about 1 in 100 of the population.
>
> Fred has his DNA is tested. It is found to match the partial DNA profile from the crime scene.
>
> The prosecution claims the DNA at the crime scene came from Fred, while the defence claims it did not.

In this scenario, *H* is the hypothesis: "The DNA at the crime scene came from Fred"

Let *E* be the evidence: "the crime scene DNA profile matches the DNA profile of Fred"

We know that if the DNA at the crime scene did NOT come from Fred, then there is a 1/100 probability of finding that its profile matches the DNA profile of Fred. In other words, we know P(*E* | *Hd*)= 1/100.

The ***prosecutor's fallacy*** is to assume that P(*E* | *Hd*) = P(*Hd* | *E*) and hence to conclude that the posterior probability of *Hd* is 1/100 This is equivalent to assuming that P(*Hp*) = 99/100 and hence to conclude that it is extremely likely the prosecutor's claim (that the DNA at the crime scene comes from Fred) is true.

In fact, despite knowing that P(*E* | *Hd*) = 1/100, we cannot make any conclusions about the posterior probability of *Hd* unless we know what the prior probability is.

Let us suppose, for example, that the crime happened on an island with 1000 people other than Fred. Then it is reasonable to assume that the prior probability P(*Hp*) is 1/1001, or equivalently that P(*Hd*)=1000/1001 since (before there is any evidence against Fred) he is no more or less likely to have left his DNA at the crime scene than anybody else.

Since the partial DNA profile from the crime scene fragment occurs in about 1 in 100 of the population, we would expect about 10 of the other 1000 people to have the same partial DNA, as shown in Figure 1. So, given the evidence *E*, we know that Fred is one of 11 people in total who could have left their DNA at the crime scene. Hence, we can conclude that

P(*Hd* | *E*) = 10/11

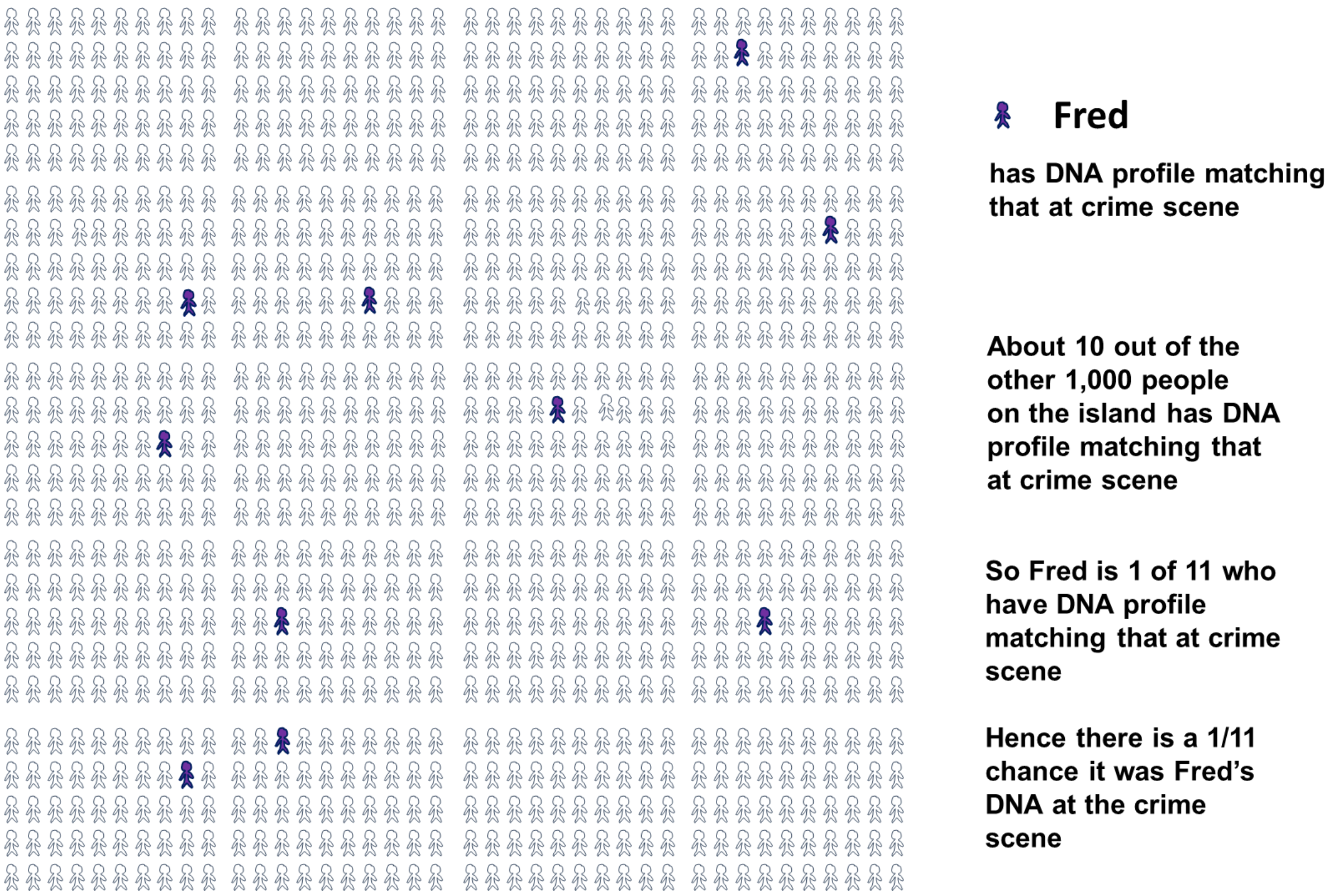

**Figure 1 Informal explanation of why P(Hd | E) = 1/11**

So, the posterior probability of the defence hypothesis is 10/11 (91%) which, of course, is very different to the 1/100 (1%) conclusion under the prosecutor's fallacy.

This was an informal calculation of the posterior probability from the prior probability P(*Hd*). and the probability P(*E* | *Hd*). The formal calculation is done using Bayes Theorem which asserts for any hypothesis *H* and evidence *E*:

$$P(H|E) = \frac{P(E|H)P(H)}{P(E)} = \frac{P(E|H)P(H)}{P(E|H)P(H) + P(E|not\ H)P(not\ H)}$$

In our scenario, P(*Hd*) = 1000/1001, P(*E* | *Hd*) = 1/100 (as already discussed). But, since *Hp* = not *Hd* we also know that P(*E* | not *Hd*) is the same as P(*E* | *Hp*) and this is equal to 1 since the probability the crime scene DNA profile matches the DNA profile of Fred will be 1 if Fred is the person who left their DNA at the crime scene. Entering these values into the Bayes Theorem equation gives us:

$$P(H_d|E) = \frac{\frac{1}{100} \times \frac{1000}{1001}}{\frac{1}{100} \times \frac{1000}{1001} + 1 \times \frac{1}{1001}} = \frac{10}{11}$$

# 2 The Likelihood ratio as a measure of probative value of evidence

Unfortunately, people without statistical training find Bayes' theorem both difficult to understand and counter-intuitive (Casscells and Graboys 1978; Cosmides and Tooby 1996). Moreover, legal professionals are also concerned that the use of Bayes requires us to assign prior probabilities. To get round this, lawyers and judges prefer to consider what they call the **probative value of the evidence** independently of the prior probability of *H*. Even if we ignore the explicit values of the prior and posterior probabilities of *Hp* and *Hd* in the above scenario, there is no doubting that the evidence *E* shifts our belief toward *Hp* being true, i.e. irrespective of the prior and posterior values it makes P(*Hp* | *E*) > P(*Hp*); hence, we say that the evidence is '**probative in favour of *H***'. The greater the shift the more probative is the evidence. Any evidence that shifts our belief toward *Hp* being false, i.e. which makes P(*Hp* | *E*) < P(*Hp*) is 'probative against *Hp*'. Evidence with 'no probative value' is evidence that does not shift our belief one way or the other, i.e. P(*Hp* | *E*) = P(*Hp*). This is important because judges normally regard such evidence as irrelevant or inadmissible (N. E. Fenton et al. 2014).

Fortunately, it turns out that it is possible to measure the probative value of evidence *E* without having to consider the prior probabilities. Specifically, ***providing that Hd is the negation is Hp*** (meaning the hypothesis states are mutually exclusive and exhaustive), we can use the Likelihood ratio (LR):

$$\frac{P(E|H_p)}{P(E|H_d)}$$

In fact, with the mutually exclusive and exhaustive assumption, we prove in the Appendix that:

1. If the LR > 1 then the evidence *E* results in an increased posterior probability of *Hp* (with higher values leading to the posterior probability getting closer to 1)
2. If the LR < 1 then the evidence *E* results in a decreased posterior probability of *Hp* (and the closer it gets to zero the closer the posterior probability gets to zero).

3. If the LR is equal to 1 then E offers no probative value since it leaves the posterior probability is unchanged.

The LR is therefore an important and valid measure of the probative value of evidence.

In our example, the DNA match evidence has a LR of 100 because

$$\frac{P(E|H_p)}{P(E|H_d)} = \frac{1/100}{1} = 100$$

If a fuller DNA profile had been found at the scene such that it occurs in, say, only 1 in a 10,000,000 people, then evidence *E* of a match in such a scenario would have a LR of 10,000,000. Evidence with such a high LR is extremely probative.

What makes the LR even more attractive as a measure of probative value of evidence is that, by Bayes Theorem

$$\frac{P(H_p|E)}{P(H_d|E)} = \frac{P(H_p)}{P(H_d)} \times \frac{P(E|H_p)}{P(E|\,H_d)}$$

And so

$$\frac{P(H_p|E)}{P(H_d|E)} = \frac{P(H_p)}{P(H_d)} \times LR$$

This is also called the 'odds version' of Bayes because the 'odds' of a hypothesis *H* is P(*H*)/P(not *H*). So, providing *Hd* = not *Hp*, it says

$$\text{Posterior odds of } H_p = \left(\text{Prior odds of } H_p\right) \times (\text{Likelihood ratio})$$

When forensic experts (and other expert witnesses), provide the likelihood ratio of their evidence they are therefore providing a universally accepted measure of the probative value of their evidence without having to express any judgement about the prior probabilities of *Hp* or *Hd*. Moreover, since the LR involves explicitly determining both P(*E*| *Hp*) and P(*E*| *Hd*) this can also help avoid the prosecutor's fallacy. Consequently, it is widely recommended for expert witnesses to report the LR when presenting their evidence. For example, use of the LR is a core recommendation in Guidelines such as (Puch-Solis et al. 2012; ENFSI 2016).

However, it is important also to recognise the following limitations of the LR:

1. ***It only works for mutually exclusive and exhaustive hypotheses***. As we have noted (and proved in the Appendix) the LR is only a valid measure of probative value if the hypotheses *Hd* and *Hd* are mutually exhaustive and exclusive (i.e. *Hd* = not *Hp*). If they are ***not*** it is possible to have LR>1 even though P(*Hd* | *E*) < P(*Hd*) and to have LR=1 even though P(*Hd* | *E*) is not equal to P(*Hd*) (N. E. Fenton et al. 2014). This crucial point is not widely known among forensic practitioners and this is concerning because, for DNA evidence it is typical ***not*** to use exhaustive hypotheses. Specifically, the hypotheses are typically:

    *Hp*: "the DNA is from the defendant".

    *Hd*: "the DNA is from a person ***unrelated*** to the defendant" (rather than the required "the DNA is NOT from the defendant").

They do this because it is statistically simpler to estimate P(*E* | *Hd*) in such cases. This standard assumption means that any LRs used for DNA match evidence are based on non-exhaustive hypotheses and are therefore not a valid measure of probative value of the prosecution hypothesis.

2. ***The LR cannot be distanced from Bayes Theorem***. A number of High Court rulings (R v T 2010; Donnelly 2005) have been extremely critical of the use of Bayes in court and, as a result, there have been attempts to claim that the LR does not involve Bayes. An indication of the extent of the confusion surrounding this can be found in one of the many responses to the RvT judgement (Aitken and many other signatories 2011). Specifically, in an otherwise excellent position statement is the extraordinary point 9 that asserts:

   > "It is regrettable that the judgment confuses the Bayesian approach with the use of Bayes' Theorem. The Bayesian approach does not necessarily involve the use of Bayes' Theorem."

   By the "Bayesian approach" the authors are specifically referring to the use of the LR, thereby implying that the use of the LR is appropriate, while the use of Bayes' Theorem may not be. But the LR is given meaning as a measure of probative value *only through Bayes theorem*. If a judge were to ask an expert witness to explain why the LR was a valid measure of probative value, the expert would be forced to reveal Bayes theorem.

3. ***Ultimately we still cannot ignore prior probabilities***. DNA evidence with a LR of 10,000,000 is obviously extremely probative. But, as impressive as that sounds, whether it is sufficient to convince you that a hypothesis is true still depends entirely on the prior probability. On an island of 1001 people (so the prior odds against *Hp* are 1 to 1000), an LR of 1,000,000 results in posterior odds of 10,000 to 1 in favour of *Hp* which may be sufficient to convince a jury that *Hp* is true. But, on an island of 60,000,001 people, the same LR of 10,000,000 results in posterior odds of 1 to 6 against *Hp*, i.e. still strongly favours the defence hypothesis.

4. ***The LR cannot always be computed by a simple formula***. We have only considered the case where there is a single piece of evidence that is directly connected to the hypothesis of interest. But, in practice, this will not generally represent reality. For example, for the DNA match evidence, once we take account of the possibility of different types of DNA collection and testing errors, then strictly speaking this involves multiple different pieces of evidence only connected to the hypothesis of interest indirectly through other unknown hypotheses (as shown in Figure 2).

5. ***The LR only tells us about the probative value of the evidence with respect to a specific hypothesis and may be highly misleading with respect to more important hypotheses***. As explained in (N. E. Fenton et al. 2014) it is possible that some evidence has little probative value on the hypothesis it is evaluated against (such as "DNA came from suspect" but may be highly probative of an even more important hypothesis such as "suspect committed the crime").

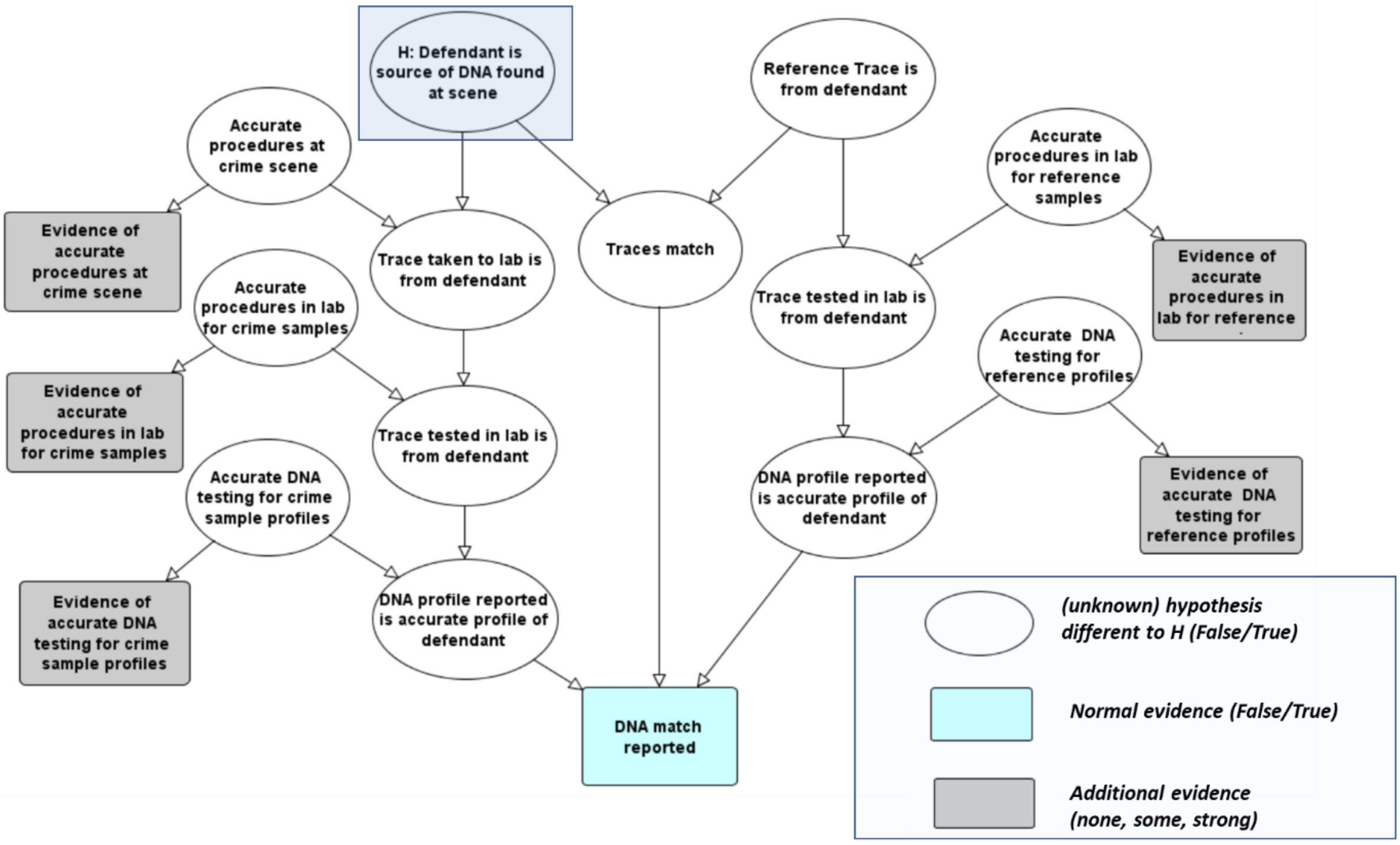

**Figure 2 Causal model (Bayesian network) for DNA match evidence**

While other concerns about the use of the likelihood ratio in legal arguments (especially with respect to its use for mixed profile DNA evidence) have been covered in (N. Fenton et al. 2020), it is the final two points above that we now focus on respectively in the next two sections.

# 3 Handling multiple pieces of evidence

Using the simple example of Fred's matching DNA profile, we can view the problem as a simple Bayesian network (BN) (N. E. Fenton and Neil 2018) as shown in **Figure 3**.

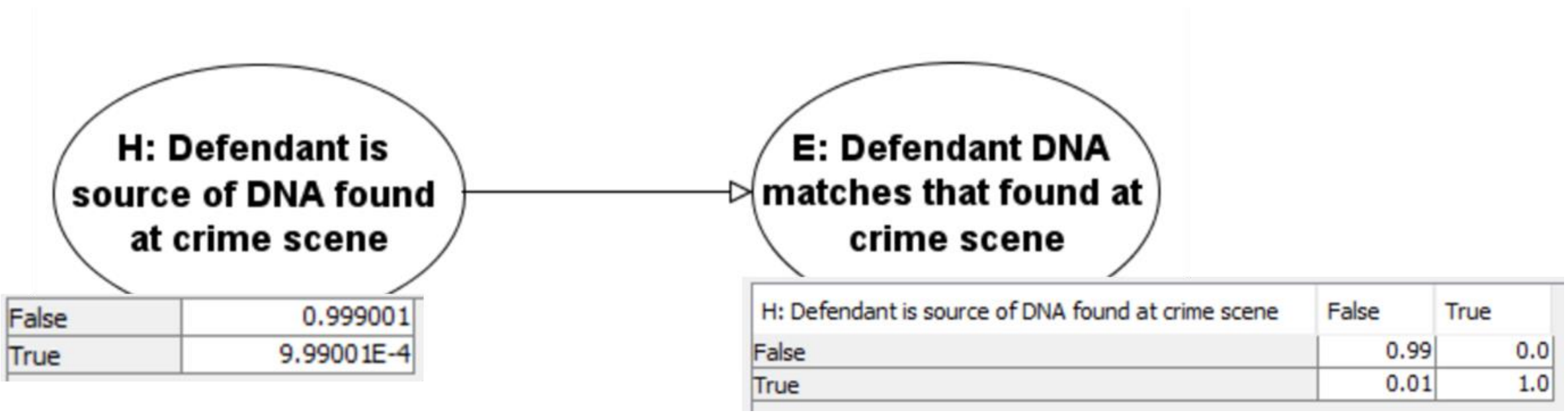

**Figure 3 Causal view of evidence, with prior probabilities shown in tables. This is a very simple example of a Bayesian Network (BN)**

A BN is directed (acyclic) graph whose nodes represent variables and for which a direct arc linking one node to another indicates our knowledge that the former directly influences or causes the latter. For each node there is an associated probability table. For nodes without any 'parent' nodes, such as node *H* in **Figure 3**, the probability table is just the prior probability of the states of the node. For nodes with parent(s), such as node *E*, the probability table expresses the probability of each state conditioned on each combination of states of the parents. So, in this simple example, the probability for E specifies the values P(*E* | *H*) and P(*E* | not *H*).

For this simple model we do not need to perform any Bayesian inference in order to calculate the LR of the evidence *E* on hypothesis *H*, because the values for P(*E* | *H*) and P(*E* | not *H*) are already directly specified in the conditional probability table for *E*.

However, what happens when there are multiple pieces of separate evidence $E_1$, …, $E_n$ and we wish to know the LR of the ***combined evidence*** on *H*?  We have to compute:

$$\frac{P(E_1, E_2, .., E_1 | H)}{P(E_1, E_2, .., E_1 | \text{not } H)}$$

In this case, whether a 'simple' calculation is possible depends on the structure of the model.

## 3.1 The simple situation when all pieces of evidence are independent conditional on *H*

It is easy to compute the LR  in the special case where each piece of evidence is independent conditional on *H* (meaning that, once we know whether *H* is true or false the probability of observing $E_i$ is independent of the probability of observing $E_j$ for any different *i* and *j*). In a BN this means that each $E_i$ corresponds to a node that is a child of *H* and there are no direct links between any of the $E_i$, as shown in Figure 4.

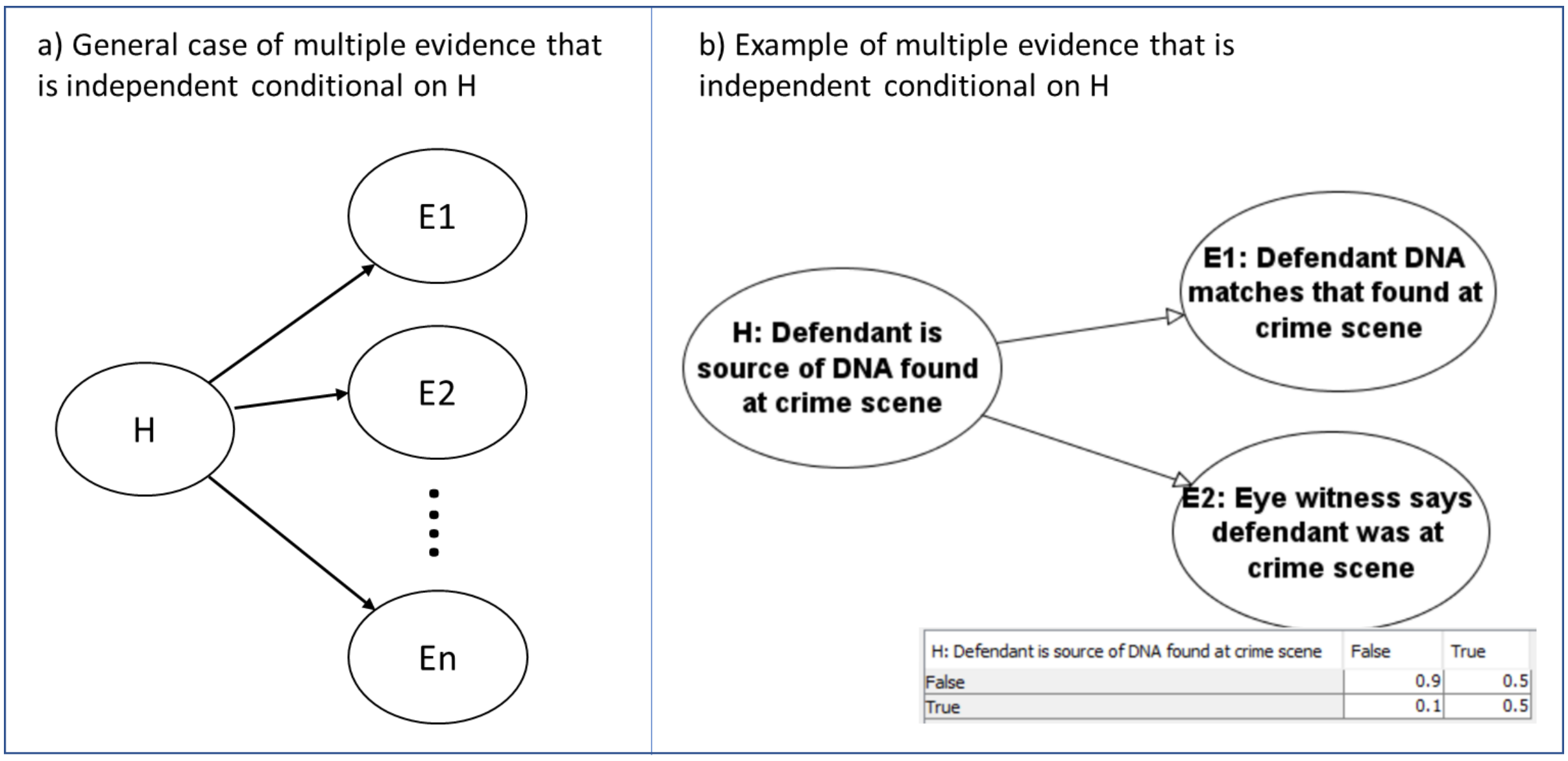

**Figure 4 Pieces of evidence that are independent (conditional on H)**

In the example of Figure 4(b), we already noted that the LR for the DNA evidence $E_1$ is 100. The LR for the eye witness evidence $E_2$ here is:

$$\frac{P(E_2|H)}{P(E_2|\text{not } H)} = \frac{0.5}{0.1} = 5$$

But in this case, because $E_1$ and $E_2$ are independent conditional on $H$, it follows that:

$$LR\ (E_1, E_2) = \frac{P(E_1, E_2|H)}{P(E_1, E_2|\text{not } H)} = \frac{P(E_1|H) \times P(E_2|H)}{P(E_1|\text{not } H) \times P(E_2|\text{not } H)}$$

$$= \frac{P(E_1|H)}{P(E_1|\text{not } H)} \times \frac{P(E_2|H)}{P(E_2|\text{not } H)} = LR\ (E_1) \times LR\ (E_2) = 100 \times 5 = 500$$

In other words, we simply multiply the separate LRs together. This extends to multiple pieces of evidence that are independent conditional on $H$, i.e. we have the following simple formula for computing the LR:

**Formula 1 (LR for independent evidence conditional on *H*)**

$$\boldsymbol{LR\ (E_1, E_2, \ldots, E_n) = LR\ (E_1) \times LR(E_2) \times \ldots LR(E_n)}$$

In the case of R V Adams (Donnelly 2005) it was exactly such a calculation that the defence presented with 3 pieces of evidence – one of which supported the prosecution hypothesis (DNA match) and two of which supported the defence hypothesis (an alibi, and a failure by the victim to identify the suspect in an identity parade).

## 3.2 The situation where there are multiple pieces of dependent evidence but only one relevant hypothesis

In Figure 5 there is a small but crucial change to the situation of Figure 4(b). In this case, it is assumed the eye witness knows the result of the DNA test before saying whether or not the defendant was at the crime scene. Because of known confirmation bias, we assume here:

- if $H$ is true and $E_1$ is true then it is more likely than in the independent case that $E_2$ will be true
- if $H$ is false and $E_1$ is true then it is more likely than in the independent case that $E_2$ will be true

It is also the case that $E_2$ is less likely to be true when $E_1$ is false for both $H$ true and false.

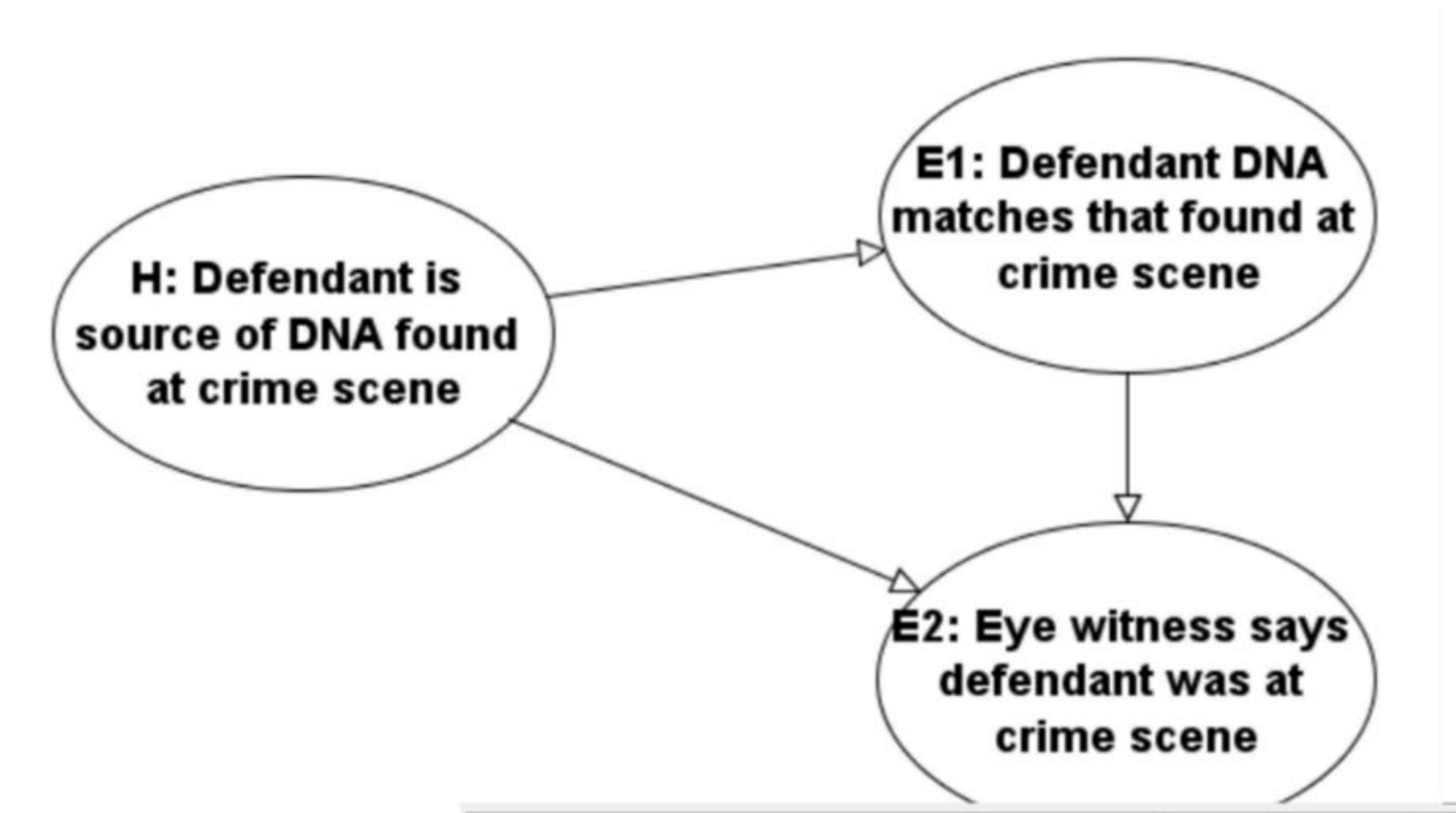

| E1: Defendant DNA matches that found at crime scene | False | | True | |
|---|---|---|---|---|
| H: Defendant is source of DNA found at crime scene | False | True | False | True |
| False | 0.99 | 0.4 | 0.7 | 0.1 |
| True | 0.01 | 0.6 | 0.3 | 0.9 |

**Figure 5 Case of dependent evidence**

In this case we cannot use the LR Formula 1 because there is no notion of a LR of $E_2$ that is independent of $E_1$. Instead we have to consider the ***dependent*** LR of $E_2$ which we will write as LR($E_2$ | $E_1$, $H$). However, we can still simply calculate the LR of the combined evidence in this case (without having to perform any Bayesian inference) using the information in the conditional probability table for $E_2$ as follows:

$$LR\ (E_1, E_2) = \frac{P(E_1, E_2|H)}{P(E_1, E_2|\text{not } H)} = \frac{P(E_1|H) \times P(E_2|E_1, H)}{P(E_1|\text{not } H) \times P(E_2|E_1, \text{not } H)}$$

$$= \frac{P(E_1|H)}{P(E_1|\text{not } H)} \times \frac{P(E_2|E_1, H)}{P(E_2|E_1, \text{not } H)} = LR\ (E_1) \times LR\ (E_2|E_1) = 100 \times \frac{0.9}{0.3} = 300$$

Note that this LR is less than that in Section 3.1 where the evidence $E_1$ and $E_2$ is independent. Hence, in this case the combined evidence is less probative.

The above formula extends to any situation where there are multiple pieces of evidence $E_1$, …, $E_n$) about the hypothesis $H$ and multiple dependencies between the evidence.

**Formula 2 (LR for dependent evidence conditional on *H*)**

**Suppose we have separate pieces of evidence *$E_i$* with possible dependencies between them. For any piece of evidence *$E_i$* let {*E*}$_i$ denote the set of 'parent' evidence nodes that *$E_i$* depends on. Then**

$$\boldsymbol{LR\ (E_1, E_2, \ldots, E_n) = LR\ (E_1|\{E\}_1) \times LR(E_2|\{E\}_2) \times \ldots LR(E_n|\{E\}_n)}$$

It is important to note that we can handle conflicting evidence using the same approach. Suppose, for example that $E_1$ is true (i.e. we have the DNA match reported) but $E_2$ is false – i.e. an eye witness (who is aware of the DNA match) says they do not recognise the defendant from the crime scene. In this case

$$LR\ (E_1, \text{not}\ E_2) = LR\ (E_1) \times LR\ (\text{not}\ E_2|E_1) = 100 \times \frac{0.1}{0.7} = 14.3$$

This compares with an LR of 100 for the DNA match evidence alone, and an LR of 55.5 for the combined $E_1$ and not $E_2$ evidence when $E_1$ and $E_2$ are independent (as in that case, from Figure 4, LR(not $E_2$) = 5/9).

## 3.3 The situation where there are multiple pieces of evidence only connected to the hypothesis of interest indirectly through other unknown hypotheses

Unfortunately, the simple methods for calculating the LR in the cases of 3.1 and 3.2 (i.e. formulas 1 and 2) completely fail for the example of **Figure 2**. In fact, it completely fails on the much simpler version shown in Figure 6.

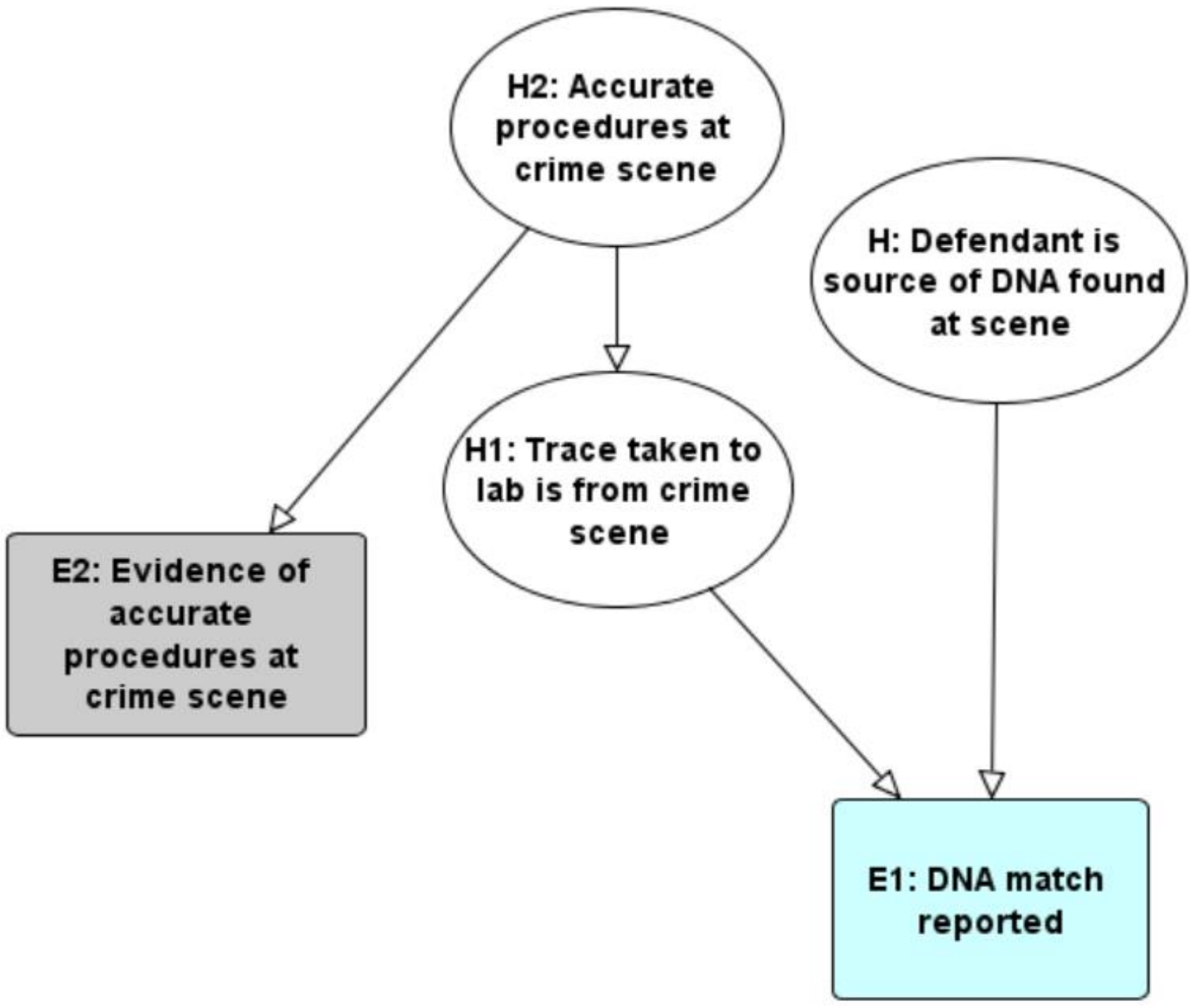

**Figure 6 DNA match evidence incorporating uncertainty about crime scene collection procedures**

In this case, because of the dependencies shown in the BN model, we have to consider the full joint probability distribution over all of the variables $E_1$, $E_2$, $H$, $H_1$, $H_2$. Using the probability chain rule and the dependency assumptions of the BN, the full joint probability distribution is:

$$P(E_1, E_2, H, H_1, H_2) = P(E_1|H, H_1) \times P(H_1|H_2) \times P(E_2|H_2) \times P(H) \times P(H_2)$$

To calculate the numerator and denominator of the LR

$$\frac{P(E_1, E_2|H)}{P(E_1, E_2|\text{not}\ H)}$$

we have to perform Bayesian updating across the whole BN which is known (in the general case) to be computationally infeasible (Cooper 1990) and which, even in this relatively simple model, is an extremely complex calculation to do from first principles. Fortunately, because there are efficient algorithms for performing inference in a wide class of BNs (Spiegelhalter and Lauritzen 1990; Pearl 1988) that have been implemented in widely available tools such as (Agena Ltd 2021; Hugin A/S 2018), we can use the results of these tools to calculate the LR. Specifically, a BN tool will calculate:

$$P(H|E_1, E_2)$$

and from this we 'recover' the LR by considering Bayes theorem as follows:

$$\frac{P(H|E_1, E_2)}{P(\text{not } H|E_1, E_2)} = \frac{P(E_1, E_2|H) \times P(H)}{P(E_1, E_2|\text{not } H) \times P(\text{not } H)} = LR \times \frac{P(H)}{P(\text{not } H)}$$

Hence, rearranging this formula we get:

$$LR = \frac{P(H|E_1, E_2)}{P(\text{not } H|E_1, E_2)} \times \frac{P(\text{not } H)}{P(H)}$$

All of the terms in this formula are automatically computed in a BN tool. Although it looks as if we have to provide a prior for P(*H*), which would defeat the objective of using the LR, ***any prior*** used will result in the same LR, and hence if we use the prior P(*H*)=0.5 the above formula becomes simply:

$$LR = \frac{P(H|E_1, E_2)}{P(\text{not } H|E_1, E_2)}$$

This is shown in Figure 7.

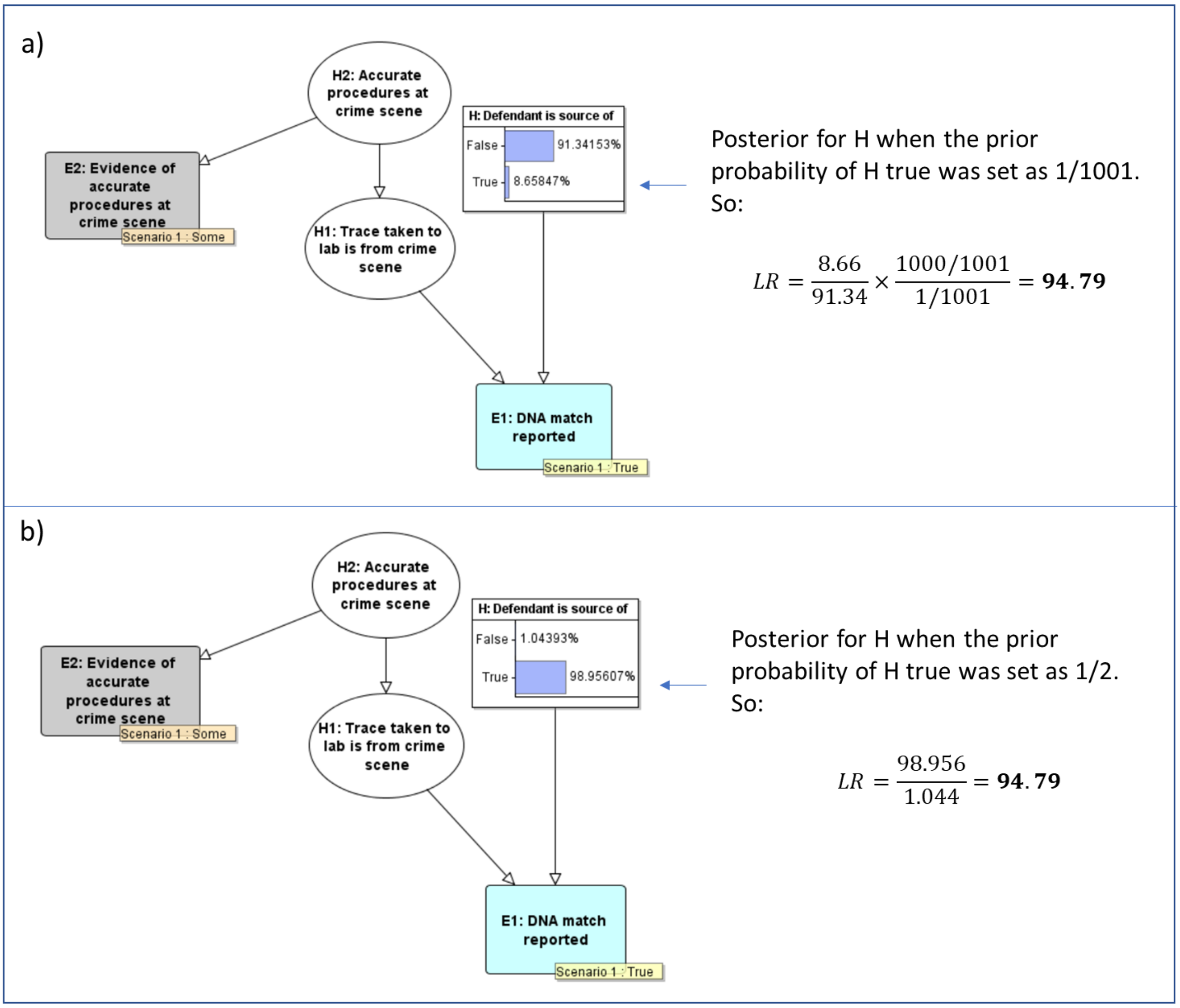

**Figure 7 The LR is determined by running the BN with the evidence and observing the posterior probability. In a) the model is run with prior P(H)=1/1001; in b) the model is run with prior P(H)=0.5. Irrespective of the prior the LR is the same: 94.79**

However, we will encounter examples where *H* has parents and hence we cannot simply set P(*H*)=0.5. In such situations we will have to use the more complex version of the formula.

Hence, in general for an arbitrary number of evidence nodes $E_1,\ldots, E_n$, irrespective of whether they are directly influenced by *H*, we have the following formulas for computing the LR from the BN:

**Formula 3. Providing *H* has no parents in the BN**

$$LR = \frac{P(H|E_1,\ldots,E_n)}{P(\text{not } H|E_1,\ldots,E_n)} \quad \text{when } P(H) = 0.5$$

**Formula 4. If *H* has parents in the BN**

$$LR = \frac{P(H|E_1,\ldots,E_n)}{P(\text{not } H|E_1,\ldots,E_n)} \times \frac{P(\text{not } H)}{P(H)}$$

Of course, one of the benefits of using a BN is that it allows us to easily consider different prior probabilities for *H*, so it would make little sense to always set P(*H*)=0.5. The BN in Figure 7(a) is providing much more information than just the LR for the specific evidence observed. It is telling us that (under the 1001 person 'island' assumption) the posterior probability that it was Fred's DNA at the crime scene is 8.66%.

In all the previous examples we have used the LR to determine the probative value of the evidence on the hypothesis of interest. However, in most situations, we also would like to know the probative value of the evidence on other (often more important) hypotheses.

# 4 Calculating the LR for related hypotheses

In all the examples so far, we have considered only a single hypothesis *H* for which we wish to know the LR for some evidence. But, in practice, we are normally more interested in knowing the probative value of the evidence on other hypotheses that may impact on *H*. For example, we have been considering the case where *H* is "DNA at crime scene comes from defendant" and where *E* is "DNA from crime scene matches that of defendant"; but while it is interesting to know the probative value of *E* on *H* it is ***more*** interesting and important to know the probative value of *E* on *H1* where *H1* is the (different) hypothesis "Defendant was at the scene of the crime" and *H2*: "Defendant committed the crime". In such cases the correct calculation of the LR for any such related hypothesis again requires us to construct a BN model and the formula for calculating the LR for any number of pieces of evidence with respect to any hypothesis is the same as formula 3 or 4 in Section 3.3.

For example, in (N. E. Fenton, Neil, and Berger 2016) we considered a (simplified version of a real) case where a suspect was accused of rape and where a tiny trace of DNA matching the suspect's DNA was found on the alleged victim's knickers. There was no dispute that the two had been together, but the defence argued that, if the DNA was from the suspect then there were several explanations for it other than through rape. The fact that there was "a DNA match" seemed to be highly probative in favour of the prosecution case as far as the lawyers (both sides), judge and jury were concerned. Yet, when considered in the full causal context the 'evidence' should have supported the defence case.

Critically, the 'evidence' is not simply '*E*: DNA match' but rather two separate pieces of evidence:

> $E_1$: DNA found on alleged victim matches DNA of defendant
>
> $E_2$: The DNA found on alleged victim was only a tiny trace

Because there was such a tiny trace it was only possible to determine the alleles on two loci and, as such, approximately 1 in 100 people would have the matching profile. So, on its own, the $E_1$ evidence has a LR of 100 in support of the prosecution hypothesis *H* "Defendant is source of DNA on victim"

However, the $E_2$ evidence is relevant not just to the (source level) hypothesis *H*, but also to the offence level hypothesis *H1*: "Defendant raped alleged victim". Specifically, if *H1* and *H* were true then finding ***only*** a tiny trace of DNA (insufficient even to determine if it came from semen or some other human tissue) is unlikely. A suitable BN for this case, with appropriate conditional probability tables, is shown in Figure 8. So, assuming the DNA is from the defendant, it assumes that there is a 10% chance that only a tiny trace would be found if the defendant raped the victim and an 80% chance if the defendant did not rape the victim.

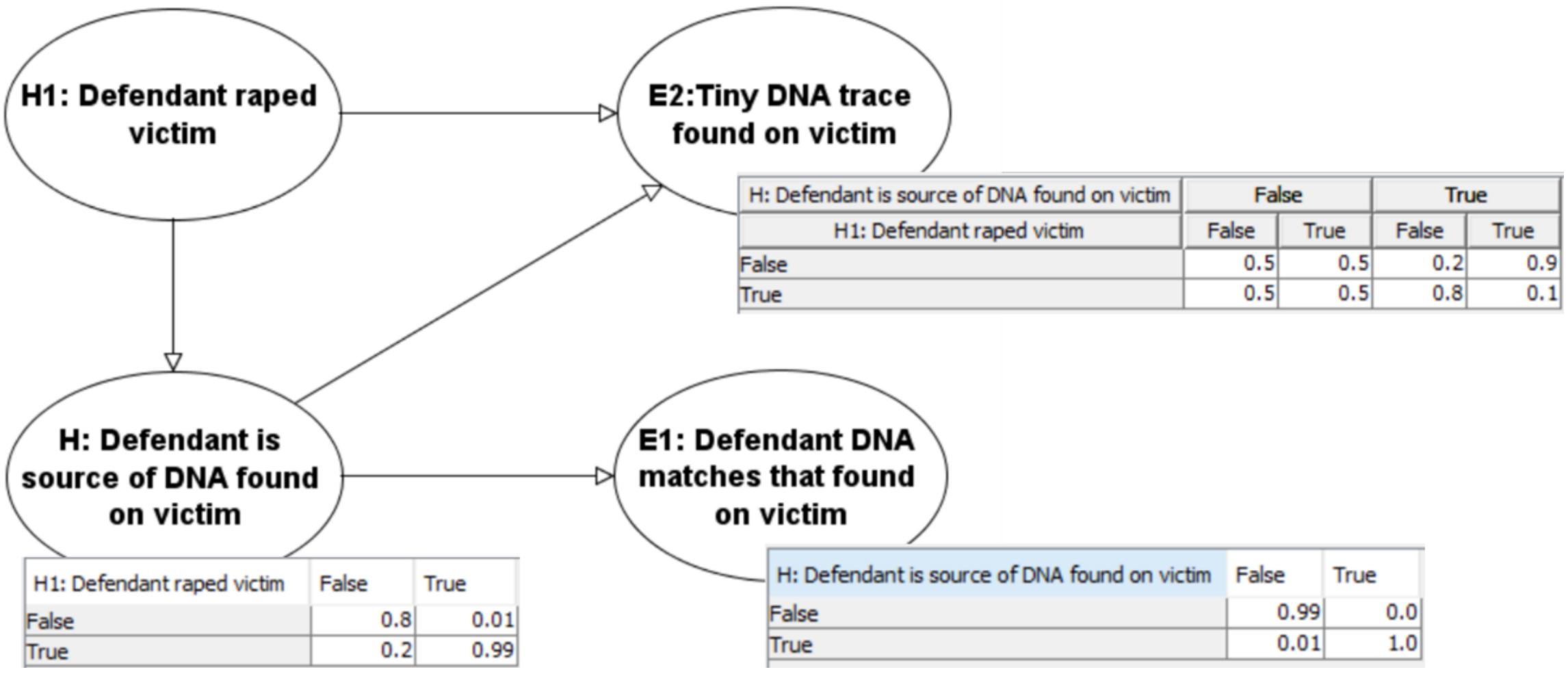

**Figure 8 BN with offence level hypothesis added**

In this situation we are much more interested in the LR of the combined evidence on *H1* rather than on *H*, but by using the BN inference we can easily calculate both (in what follows we assume a 50% prior for *H1* being true so that we can use the simpler the LR calculation in Formula 3 of Section 3.3). But first it is important to see the separate impact of $E_1$ alone as shown in Figure 9.

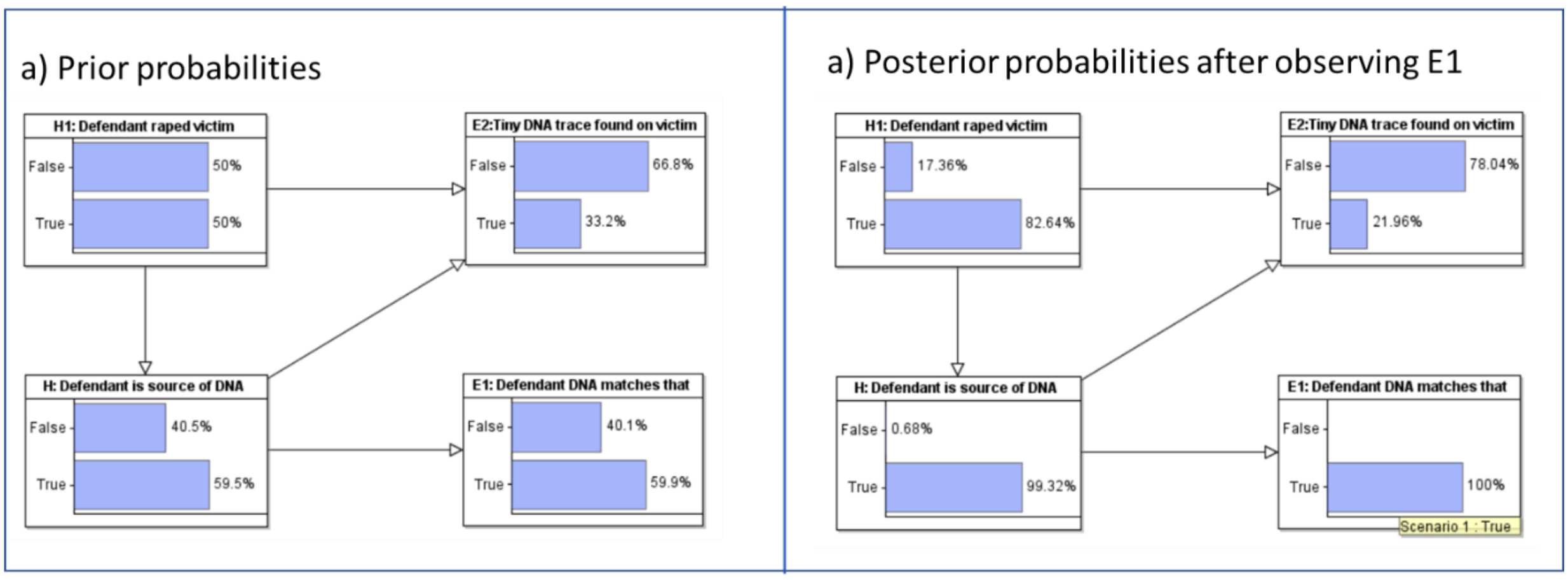

**Figure 9 Impact of DNA match evidence E1. Prior (a) and posterior (b). The probability both H and H1 true increases so the evidence clearly supports the prosecution case.**

The evidence is clearly probative in favour of both the prosecution hypotheses *H1* and also *H* and. We already know the LR of $E_1$ on *H* is 100. Because the posterior probability of *H1* true is 82.64% and *H1* false is 17.36%, the LR of $E_1$ on *H1* is 82.64/17.36 = 4.8.

However, things are very different when we consider the impact of the combined evidence as shown in Figure 10.

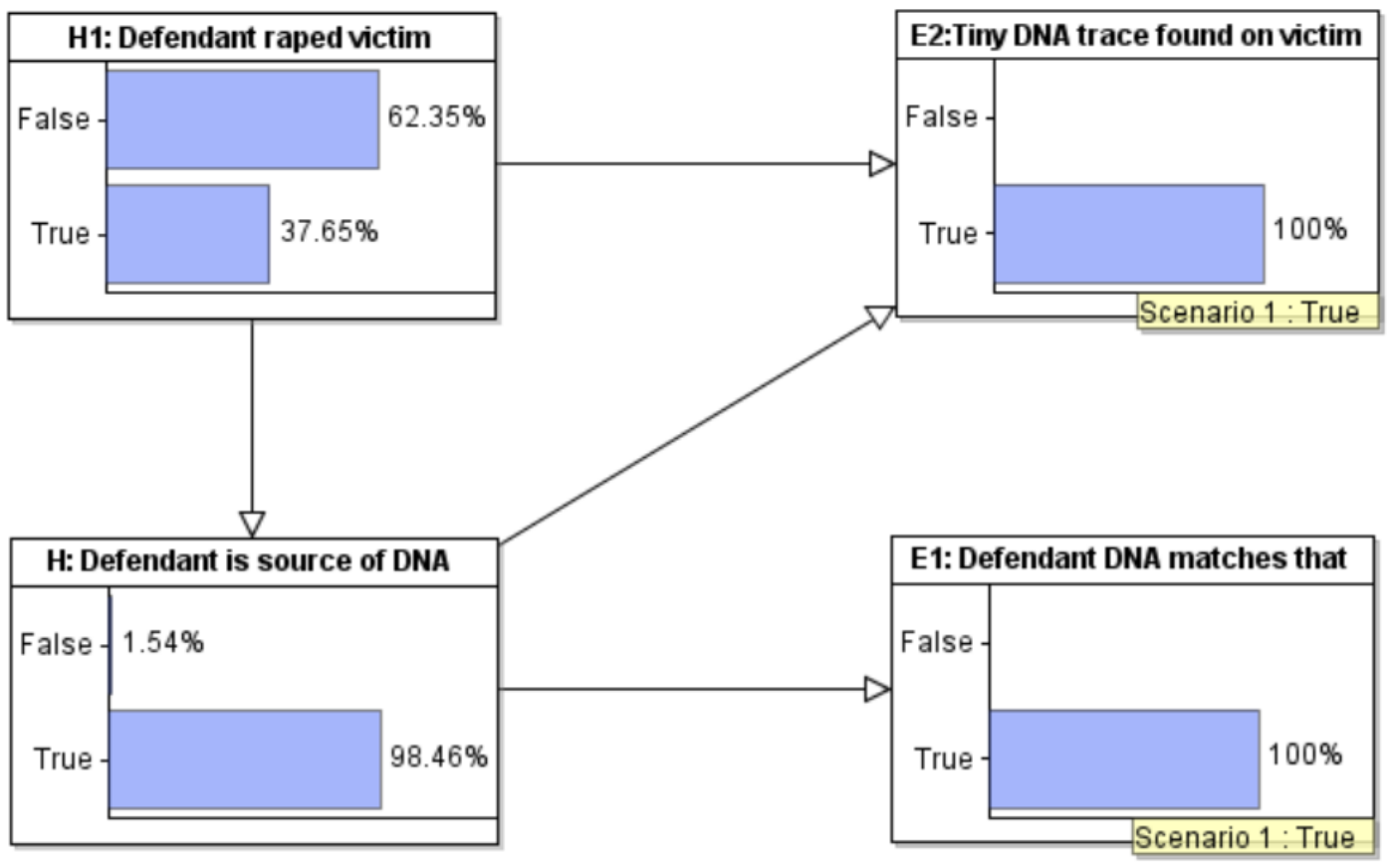

**Figure 10 Impact of the combined evidence**

While the LR of the combined evidence on *H* is still highly probative in favour of *H*, the LR of the combined evidence on *H1* is 37.65/62.35 = 0.6. Being less than 1 this means the combined evidence supports the defence hypothesis (*H1* false) rather than the prosecution hypothesis (*H1* is true). The prior probability of *H1* dropped from 50% to 37.65%.

A more comprehensive example is shown in Figure 11. In this case the suspect is accused of a crime that took place in a pub. The pub was known to have had 100 people in it at the time of the crime and the suspect admits to having been there earlier but claims that he was in his home nearby at the time of the crime. This places him in an 'extended crime scene' (N. E. Fenton et al. 2019) estimated to contain 1000 people, meaning the prior probability the suspect was in the pub at the time of the crime is 1/10, the prior probability of of *H1* (the suspect was the offender) is 1/1000 and the prior probability that DNA found in the pub is from the suspect is 0.117. As in our previous example, we have evidence *E* that DNA found at the crime scene matches that of the suspect. In this case the DNA profile is found in only 1 in 10,000 people. However, we assume there is some possibility the sample is corrupted and, also, that there may have been a lab error.

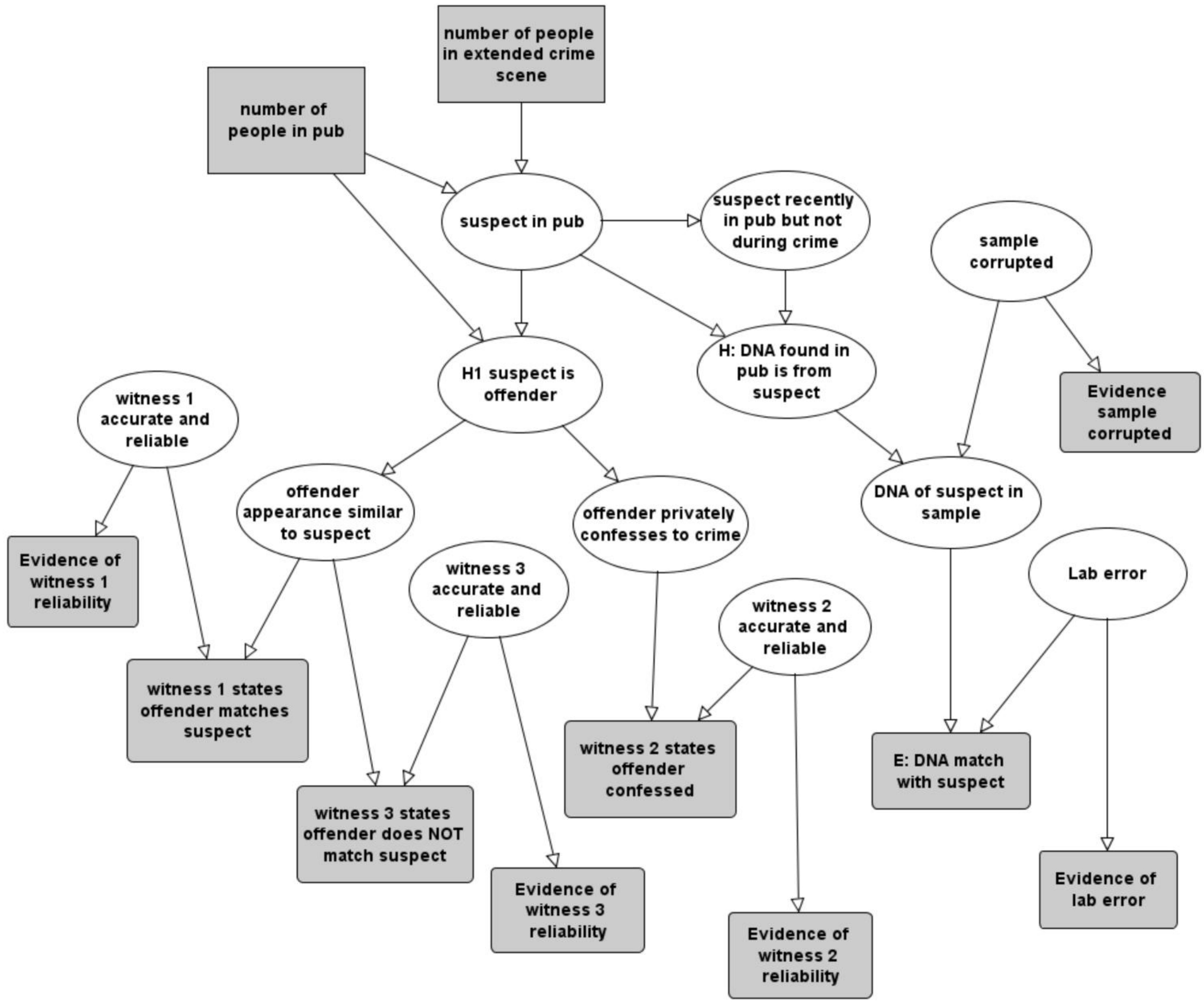

**Figure 11 Comprehensive case example. Square grey nodes represent (possible) evidence, while round white nodes are all uncertain hypotheses.**

Suppose, initially we get the DNA match evidence supported by evidence that there was no lab error and the sample was not corrupted. The impact of this combined evidence is shown in Figure 12.

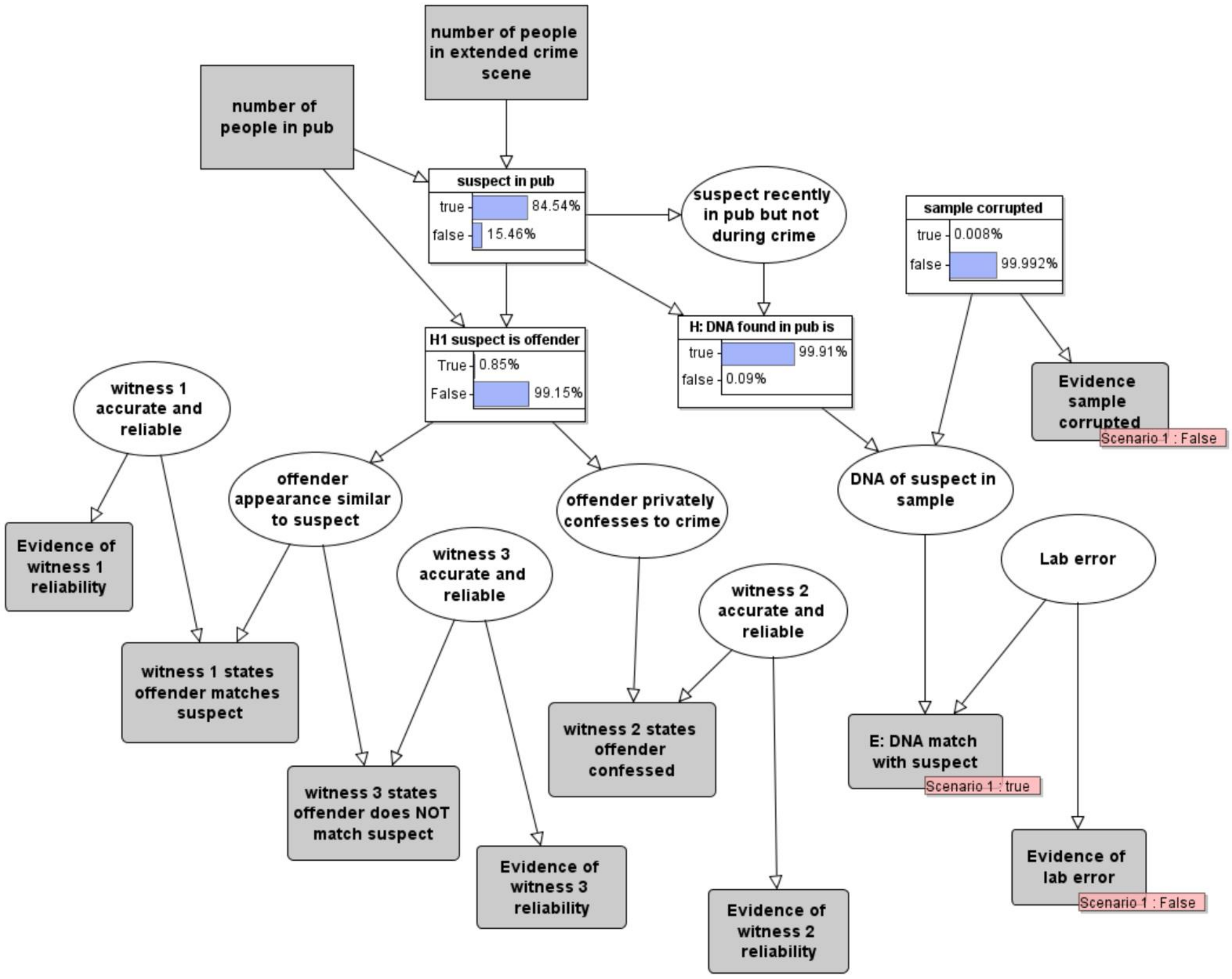

**Figure 12 Impact of the DNA match evidence**

Note the combined evidence is highly probative for the hypothesis *H* – that the DNA found at the crime scene came from the suspect. As the prior probability was 11.7% and the posterior probability is now 99.909% the LR for the combined evidence on *H* is (using Formula 3):

$$\frac{99.909}{0.091}\times\frac{88.3}{11.7}=8286$$

which is clearly highly probative. But its impact on the offence level hypothesis *H1* is much less, since the prior of 1/1000 has only moved to 0.85%. So, the LR of the combined DNA evidence on *H1* is:

$$\frac{0.85}{99.15}\times\frac{999/1000}{1/1000}=8.56$$

However, suppose an eye witness for the prosecution (witness 1) testifies that the suspect is similar in appearance to the offender and that another witness for the prosecution (witness 3) testifies that the suspect admitted to being the offender while in jail on remand. Suppose there is evidence that these witness are ‘possibly reliable’. Then the overall impact of all the combined evidence is shown in Figure 13.

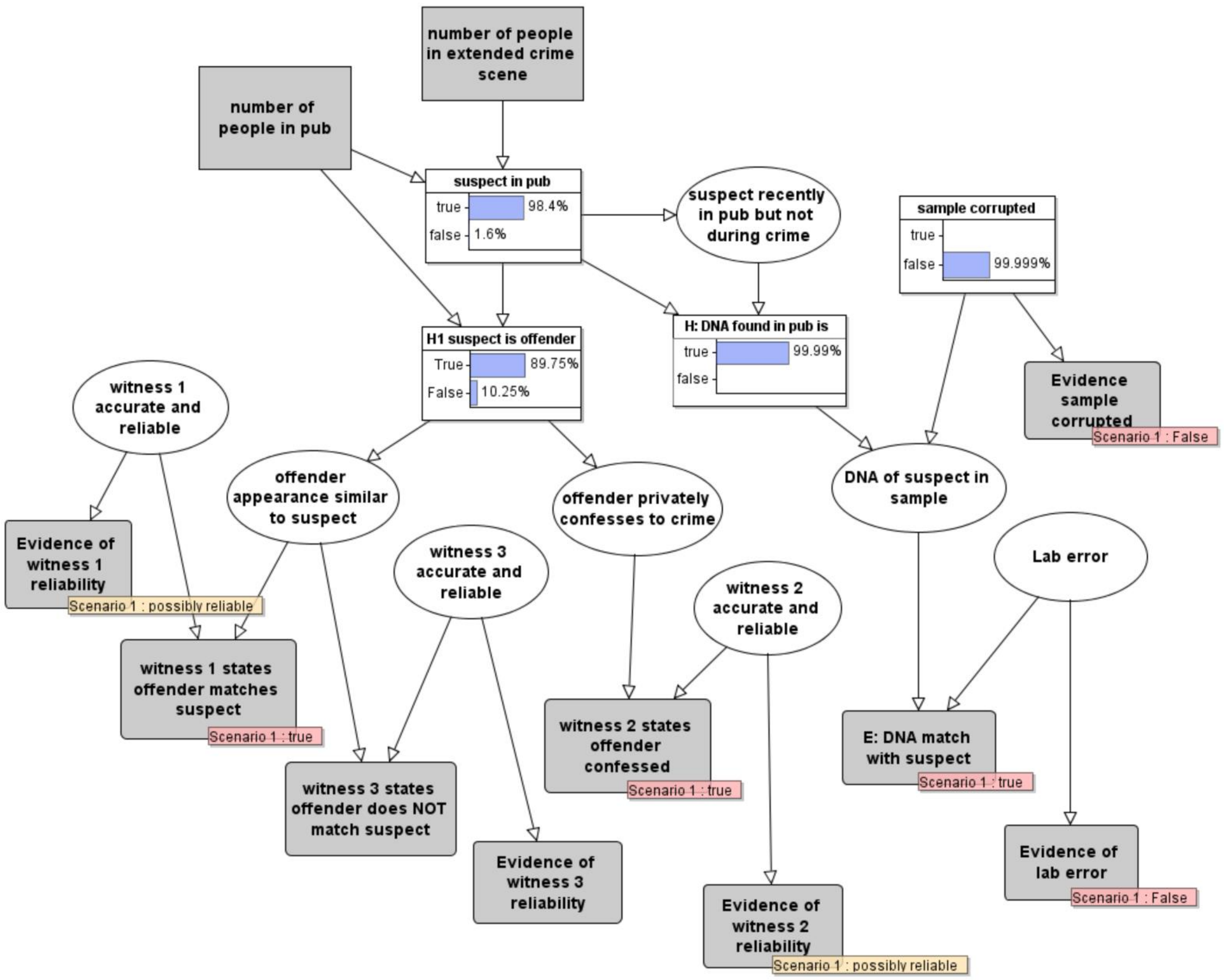

**Figure 13 Prosecution witness evidence added**

The LR of this combined evidence on the offence level hypothesis *H1* is now:

$$\frac{89.75}{10.25} \times \frac{999/1000}{1/1000} = 8747$$

This is highly probative in favour of *H1* being true and indeed the posterior probability of 89.75 may even be considered sufficient to find the suspect guilty.

However, suppose another eye witness who is considered very reliable testifies that the suspect is NOT similar in appearance to the offender. Then the overall impact of the combined evidence now is shown in Figure 14.

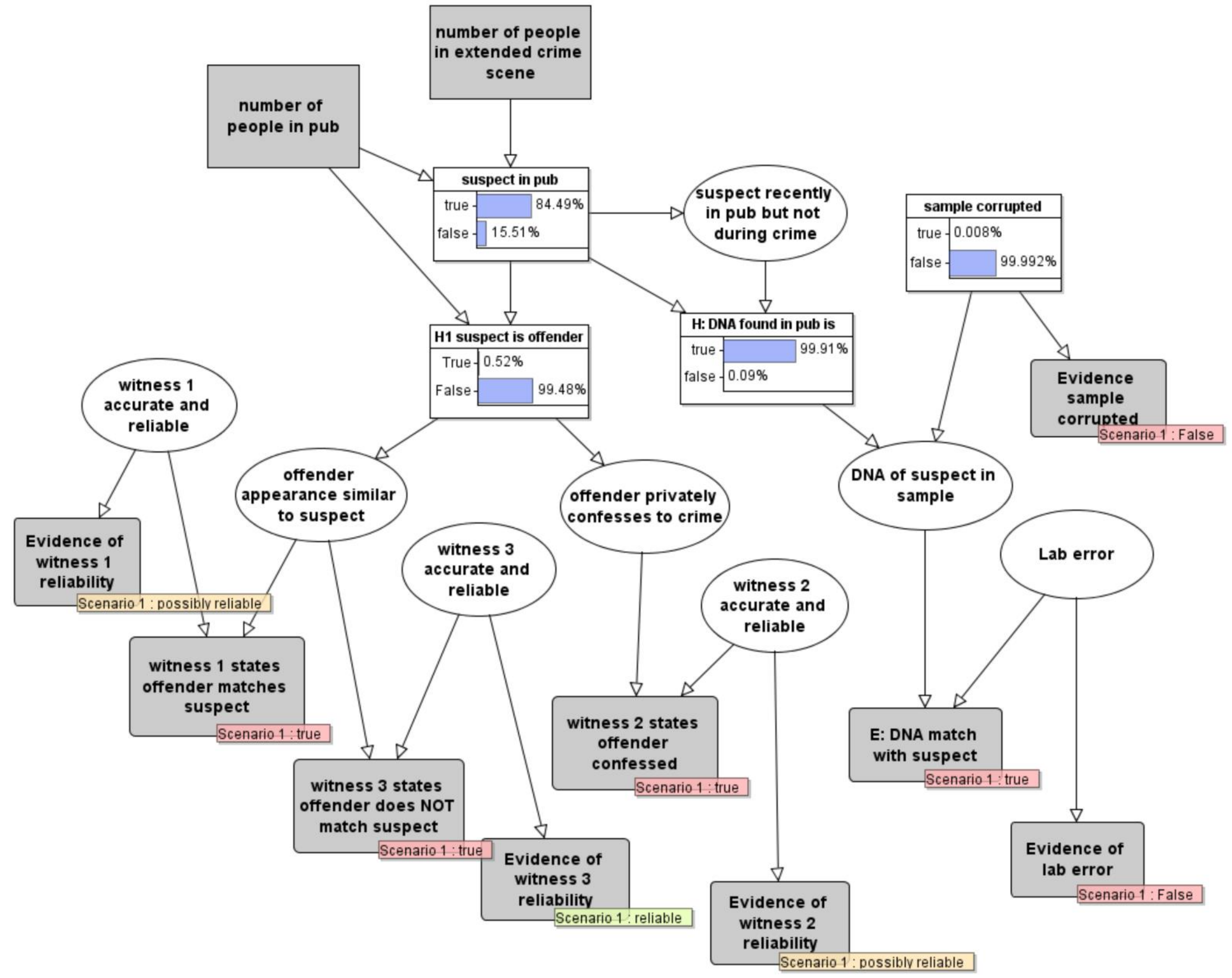

**Figure 14 New reliable eye witness testifies that suspect was NOT similar in appearance to the offender**

The LR of all the combined evidence on *H1* is now:

$$\frac{0.052}{99.48} \times \frac{999/1000}{1/1000} = 0.522$$

In other words, despite all the DNA evidence and the prosecution witness evidence, the addition of this one reliable eyewitness means the combined evidence supports the defence case rather than the prosecution.

# 5 Can relevant hypotheses be conditioned on evidence?

Although we have seen examples of evidence conditioned on other evidence, we have generally avoided examples in which hypotheses are conditioned on evidence, i.e. examples in which an evidence node in the BN model is a parent of a hypothesis node. This is because, generally, evidence is something that follows both causally and temporally from a hypothesis; If the suspect committed the crime or was at the crime scene then this may result in finding evidence like forensic traces left by the suspect or sightings by eyewitnesses. No such evidence can ***cause*** the suspect to have committed/not committed the crime or to be/not be at the crime scene.

However, it is often assumed that there are two types of evidence that ***are*** necessary (causal) requirements for a crime to be committed: *motive* and *opportunity*.

We contend that 'suspect had motive' is a hypothesis (never known for certain) about which we may or may not find evidence as shown in Figure 15. Hence, it should not be considered as direct evidence.

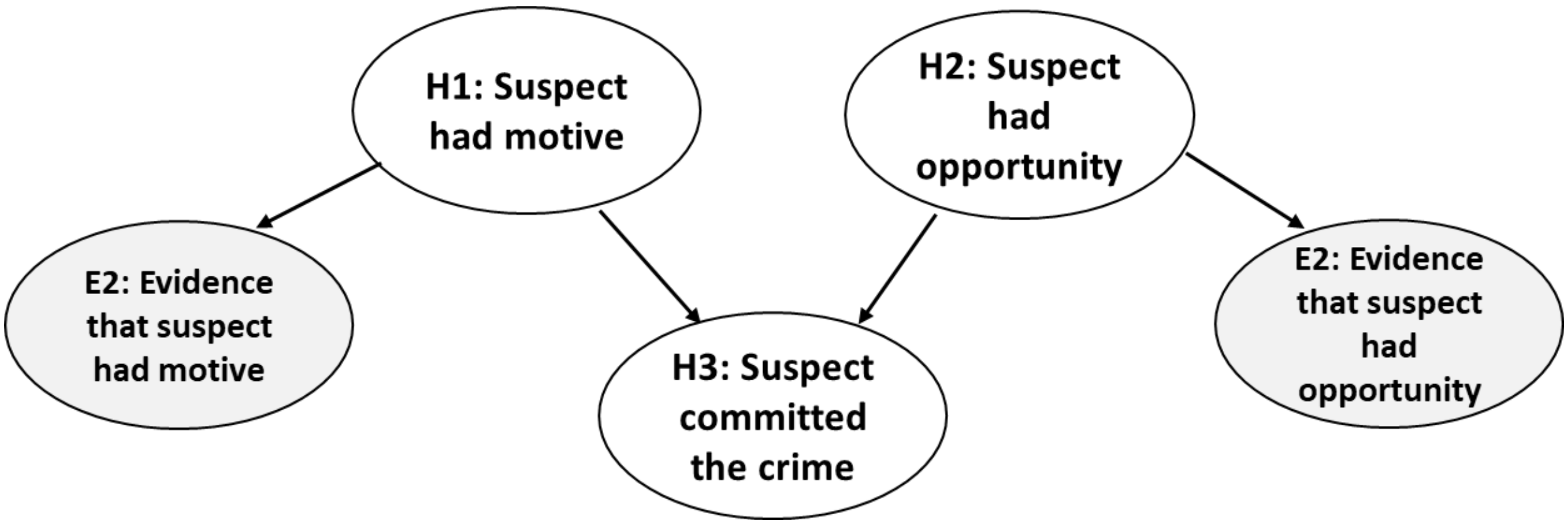

**Figure 15 Dealing with motive and opportunity**

The same applies to opportunity, which usually is just the hypothesis "the suspect was at the crime scene". However, note that in the example of Figure 11, there appears to be an exception to the rule with the evidence nodes: 'number of people in pub' and 'number of people at extended crime scene'. This follows the approach to defining the 'opportunity prior' (N. E. Fenton et al. 2019), whereby it is possible for certain types of crime to establish both a prior for being at the crime and a prior for committing the crime based purely on information about the number of people who were within the 'extended crime scene'; this is the smallest area – in distance and time (in which the suspect was known to be present) from the actual crime scene and time. For example, if there is no dispute that the suspect was in the pub at the time of the crime there and if there were 100 people in total in the pub then the 'opportunity prior' of 1/100 in this case is equal to the prior probability of guilt. In the example it was not known if the suspect was in the pub, but it was known he was in an extended crime scene with 1000 people. Hence, the prior probability of guilt is 1/1000 and the prior probability of being at the crime scene is 1/10.

In fact, the use of evidence nodes that are direct parents of hypothesis nodes in Figure 11 is just a simplification. Strictly speaking the 'number of people in pub' and 'number of people at extended crime scene' are both hypothesis nodes which should have evidence nodes as children.

In practice, whether we do choose to condition hypothesis nodes directly on evidence or not, we can still compute the LR using Formula 4.

# 6 Conclusions

While the LR is a valid and useful measure of probative value of evidence for a hypothesis with mutually exclusive and exhaustive states, its computation is not always as simple as many seem to believe. If we have to consider the LR for any type of combined evidence on a hypothesis for which at least one piece of evidence is not directly related, then the calculation of the LR requires us to build a BN. Specifically, the BN defines the dependency structure of the hypotheses and evidence and the conditional probability tables of the BN define the ‘localised’ LRs. To compute the LR of the combined evidence on a hypothesis in the BN we run the model in a BN tool with the evidence entered (any BN tool will do the necessary Bayesian updating) and observe the posterior probability for the hypothesis of interest. If its prior probability was set as 0.5 (which is always possible if the node has no parents) then the LR is simply the posterior probability *H* is true divided by the posterior probability *H* is false. If the hypothesis of interest has parents then it may not be possible to set its prior probability as 0.5, but in this case the LR is: (the posterior probability *H* is true divided by the posterior probability *H* is false) times the prior probability *H* is false divided by the prior probability *H* is true).

## Appendix 1 Bayes' Theorem proof and the probative value of evidence for mutually exclusive hypotheses

When prosecution and defence hypotheses are mutually exclusive, a LR of greater than one supports the prosecution hypothesis and a LR of less than one supports the defence hypothesis. Hence, the LR has a simple interpretation for the probative value of the evidence under these assumptions, and the proof is as follows:

In order to prove this important property of the LR, we need Bayes' theorem

Bayes' Theorem tells us that:

$$P(H \mid E) = \frac{P(E \mid H)P(H)}{P(E)}$$

By applying Bayes' theorem to both $H_p$ and $H_d$ we get the equivalent form of Bayes (called the 'odds' version):

$$\frac{P(H_P \mid E)}{P(H_D \mid E)} = \frac{P(E \mid H_P)}{P(E \mid H_D)} \times \frac{P(H_P)}{P(H_D)}$$

In this version the term

$$\frac{P(E \mid H_P)}{P(E \mid H_D)}$$

is the likelihood ratio (LR) – it is simply the prosecution likelihood divided by the defence likelihood.

The term

$$\frac{P(H_P)}{P(H_D)}$$

represents the 'prior odds' – the relative prior belief in the prosecution hypothesis over the defence hypothesis.

The term

$$\frac{P(H_P \mid E)}{P(H_D \mid E)}$$

represents the revised 'posterior odds' – the relative (posterior) belief in the prosecution hypothesis over the defence hypothesis having observed the evidence E.

Most texts that attempt to explain the impact of the LR on the probative value of *E* use an argument based on the relative 'odds' of the hypotheses. The formula tells us that whatever our prior odds were in favour of the prosecution hypothesis, the posterior odds are the result of multiplying the prior odds by the LR. Hence, when the prosecution likelihood is greater than the defence likelihood the posterior odds in favour of the prosecution hypothesis must increase.

However, this argument it is unnecessarily confusing, because not only does it hide the assumption that the hypotheses need to be mutually exclusive for it to work, but it also fails to tell us clearly what we most need to know: namely that for the evidence *E* to 'support' the hypothesis $H_p$ it is necessary that the posterior probability of $H_p$, i.e. $P(H_p \mid E)$ is greater than the prior probability $P(H_p)$ in other words our belief in $H_p$ being true increases after we observe E.

What follows is a proof that $P(H_p \mid E) > P(H_p)$ when the LR is greater than 1:

From Bayes' Theorem:

$$\frac{P(H_P \mid E)}{P(H_D \mid E)} = \frac{P(E \mid H_P)}{P(E \mid H_D)} \times \frac{P(H_P)}{P(H_D)}$$

But since the LR > 1 it follows that:

$$\frac{P(H_P \mid E)}{P(H_D \mid E)} > \frac{P(H_P)}{P(H_D)}$$

But because $H_d$ = not $H_p$ we know that

$P(H_D) = 1 - P(H_P)$ and $P(H_D \mid E) = 1 - P(H_P \mid E)$

Hence, substituting these into the above inequality equation we get:

$$\frac{P(H_P \mid E)}{1 - P(H_P \mid E)} > \frac{P(H_P)}{1 - P(H_p)}$$
$$\Rightarrow P(H_P \mid E)(1 - P(H_p)) > P(H_P)(1 - P(H_P \mid E)$$
$$\Rightarrow P(H_P \mid E) - P(H_p)P(H_P \mid E) > P(H_P) - P(H_P)P(H_P \mid E)$$
$$\Rightarrow P(H_P \mid E) > P(H_P)$$

## Appendix 2. Neutral evidence

First we prove that evidence *E* is neutral when the LR is 1 and when the prosecution and defence hypotheses are mutually exclusive.

Since the LR is 1 we know that $P(E \mid H_p) = P(E \mid H_d)$

$$
\begin{aligned}
P(H_P \mid E) &= \frac{P(E \mid H_P)P(H_P)}{P(E \mid H_P)P(H_P) + P(E \mid H_D)P(H_D)} \\
&= \frac{P(E \mid H_P)P(H_P)}{P(E \mid H_P)P(H_P) + P(E \mid H_P)P(H_D)} \text{ since } P(E \mid H_P) = P(E \mid H_D) \\
&= \frac{P(E \mid H_P)P(H_P)}{P(E \mid H_P)(P(H_P) + P(H_D))} \text{ since } P(E \mid H_P) = P(E \mid H_D) \\
&= \frac{P(E \mid H_P)P(H_P)}{P(E \mid H_P)} \text{ since } P(H_P) + P(H_D) = 1 \text{ as } P(H_P), P(H_D) \text{ mutually exclusive and exhaustive} \\
&= P(H_P)
\end{aligned}
$$

What happens when the LR =1 but $H_p$ and $H_d$ are not mutually exclusive? From the odds version of Bayes' we know that

$$\frac{P(H_P \mid E)}{P(H_D \mid E)} = \frac{P(H_P)}{P(H_D)}$$

So all we can *actually* conclude is that the ratio of the posterior probabilities of $H_p$ and $H_d$ is equal to the ratio of the prior probabilities.